\documentstyle[aps,prb,preprint,tighten]{revtex}

\begin{document}
\draft
\preprint{STCS preprint}
\title{Relation of Extended Van Hove Singularities \\
to High-Temperature Superconductivity \\
within Strong-Coupling Theory}
\author{R. J. Radtke}
\address{Department of Physics and the James Franck Institute, \\
The University of Chicago, Chicago, Illinois  60637}
\author{M. R. Norman}
\address{Materials Science Division,
Argonne National Laboratory, Argonne, Illinois  60439}
\maketitle

\begin{abstract}
Recent angle-resolved photoemission (ARPES) experiments have indicated that
the electronic dispersion in some of the cuprates possesses an extended saddle
point near the Fermi level which gives rise to a density of states
that diverges like a power law instead of the weaker logarithmic divergence
usually considered.
We investigate whether this strong singularity can give rise to
high transition temperatures by computing the critical temperature $T_c$
and isotope effect coefficient $\alpha$
within a strong-coupling Eliashberg theory which accounts for the
full energy variation of the density of states.
Using band structures extracted from ARPES measurements,
we demonstrate that, while the weak-coupling solutions suggest
a strong influence of the strength of the Van Hove singularity on $T_c$
and $\alpha$,
strong-coupling solutions show less sensitivity to the singularity
strength and do not support the hypothesis that band structure effects alone
can account for either the large $T_c$'s or the different $T_c$'s within the
copper oxide family.
This conclusion is supported when our results are plotted as a
function of the physically relevant self-consistent coupling constant,
which show universal behavior at very strong coupling.
\end{abstract}

\bigskip


\narrowtext

On general grounds, we know that low-dimensional features in the
electronic dispersion lead to singularities in the density of states,
and we also expect that the superconducting critical temperature $T_c$
will be enhanced when the Fermi level lies near one of these singularities.
Early work along these lines focused on the A15 superconductors, where
the low-dimensional features are one-dimensional chains,
and calculations in both BCS \cite{FriedelA15,CarbotteWC}
and Eliashberg \cite{Horsch,CarbotteSC,Pickett} formalisms demonstrated
a small but significant enhancement of $T_c$.
Current interest is in the family of high-temperature cuprate superconductors,
where the low-dimensional structures are the $CuO_2$ planes.
Energy dispersions from photoemission measurements and LDA band structure
calculations led a number of workers to suggest that the density of
states in these compounds was logarithmically divergent and that this
divergence, in addition to a fairly typical pairing boson, could account
for the large critical temperatures, the
anomalous isotope effect, and the specific heat.
\cite{Hirsch,FriedelHTSC,Tsuei,Mark}
This picture of the origin of high-temperature superconductivity has
become known as the ``Van Hove scenario.''
While a logarithmic singularity is appropriate for
$La_{2-x}Sr_xCuO_4$ (LSCO), recent photoemission measurements on
$YBa_2Cu_3O_{6.9}$ (YBCO), $YBa_2Cu_4O_8$, and $Bi_2Sr_2CaCu_2O_8$
(BSCCO) suggest that the density of states in these materials has a much
stronger power law divergence. \cite{Campuzano,Abrikosov,Shen}
Weak-coupling calculations suggest that the enhancement of $T_c$ due
to these power law singularities would be much larger than that
for the logarithmic singularity and may account for the larger $T_c$'s
observed in these compounds. \cite{Abrikosov}

In this paper, we examine this scenario
by using tight-binding fits to photoemission data
as input to a strong-coupling Eliashberg $T_c$ calculation in
order to determine whether power law Van Hove singularities significantly
enhance $T_c$'s over those given by logarithmic ones.
We also quantify the $T_c$ enhancement as a function of singularity strength
relative to a flat density of states.
We find that the dramatic increase in $T_c$ due to the Van Hove singularity
in weak coupling theory is moderated considerably by strong-coupling
effects and that the enhancement of $T_c$ by a power law
singularity over a logarithmic one is also reduced somewhat by
strong-coupling effects.
We suggest several possible extensions of this work, though, which
should be investigated before drawing a negative conclusion
concerning the Van Hove scenario for the cuprates.

For computing the critical temperatures,
we use the standard mean field formalism \cite{AM} in which the electron
self-energy is solved self-consistently from the single-exchange graph
generalized to include the energy dependence of the density of states.
\cite{CarbotteSC,Pickett,Radtke}
For simplicity, we take the pairing interaction to be an Einstein phonon
with frequency $\Omega_0$ = 500 K (43 meV), and we assume that the
electron-phonon matrix element $g$ is independent of wave vector.
The equations for the electron self-energy in Matsubara space
$\Sigma (i \omega_n)$
can then be written in the Nambu matrix notation \cite{AM} as
\begin{eqnarray}
\Sigma (i \omega_n)
  & = & -T \sum_{m} g^2 D (i\omega_n - i\omega_m)
          \int dE \, N(E)
          \, \tau_3 G (E, i\omega_m) \tau_3
  \label{eq:El}
\end{eqnarray}
where $G (E, i \omega_n)$ is the electron
Green's function which satisfies the Dyson equation
\begin{eqnarray}
G^{-1} (E, i \omega_n)
  =  G_0^{-1} (E, i \omega_n) - \Sigma (i \omega_n)
  \label{eq:GF}
\end{eqnarray}
with $G_0 (E, i \omega_n)$ being the bare propagator.
In these expressions, $T$ is the temperature, $g^2$ is the electron-phonon
coupling constant, $D$ is the phonon propagator, $N(E)$ is the single-spin
density of states per unit cell, $\tau_3$ is the third Pauli matrix, and
$\omega_n = (2n + 1)\pi T$ are Fermionic Matsubara frequencies.
Throughout this paper, we set $\hbar = k_B = 1$.
We solve these self-energy equations for
all components of the Nambu self-energy ($\tau_0$, $\tau_1$, and $\tau_3$)
\cite{AM} and at fixed band filling $n$ determined by
\begin{equation}
n = 1 + \sum_{n} \, e^{-i\omega_n 0+} \, \int dE \, N(E) \,
     {\rm Tr} \left[ \tau_3 G(E, i\omega_n) \right] . \label{eq:bf}
\end{equation}
We fix $n$ in order to properly account for the loss of particle-hole
symmetry in the interacting system, but we note that, due to the uncertainties
in the parameterization of the band structures from ARPES data described
below, the absolute values of $n$ should not be taken too seriously.
Note that this approach includes the effects of inelastic pair-breaking, which
is important for obtaining reasonable $T_c$ estimates.
We linearize these equations in the gap function and solve the resulting
eigenvalue problem for $T_c$.
We also extract isotope effect coefficients $\alpha$ from the relation
\begin{equation}
\alpha = \left. \frac{1}{2} \, \frac{d \ln T_c}{d \ln \Omega_0} \right|_\lambda
\end{equation}
by changing the Einstein phonon frequency $\Omega_0$ a small amount
and adjusting $g^2$ so that $\lambda = 2 g^2 / \Omega_0 W$, $W$ the
band width, is constant.

In order to complete the specification of the problem, we must discuss
the band structures used to determine $N(E)$
and the choice of coupling constants.
We obtain the electronic dispersion $E_{\bf k}$ by fitting
to a two-dimensional, tight-binding form
$E_{\bf k} = \sum_{i = 0}^{5} t_i \eta_i ({\bf k})$
with the basis functions $\eta_i ({\bf k})$ listed in Tab. 1 and
the parameters $t_i$ in Tab. 2.
Representative dispersions for LSCO and BSCCO are shown in Fig. 1.

Our fitted band structures are obtained from different procedures, which
we will now describe.
For the LSCO case, a simple two-parameter fit was done to an LDA band structure
calculation with a 2 eV band width.
For the YBCO case, we perform three fits to the data.
The first fit was done by assuming a 2 eV band width
as in LSCO and adding a third tight-binding term to force a $k^4$ dependence
along the $k_x$ direction around the $(\pi,0)$ point.
This procedure gives an extended saddle point and so approximates the
photoemission data.\cite{Campuzano,Abrikosov}
A second fit was done by relaxing the third parameter to best
fit all the data along $(0,0)$ to $(\pi,0)$.
This method splits the extended singularity at $(\pi,0)$ into two
nearby logarithmic singularities (bifurcated saddle points) as seen in LDA
calculations, which are related to the displacement of the planar oxygens along
the c axis of the crystal. \cite{Andersen}
(The data are more consistent with an extended
saddle point about $(\pi,0)$ than bifurcated saddle points.  The
flatness of the experimental bands is almost certainly a correlation
effect, which can only be approximately simulated by the tight-binding fit.)
The third fit assumed a 1 eV band width and gives a better fit to
the data along the $(\pi,0)$ to $(\pi,\pi)$ direction, which are
consistent with a quadratic dispersion about $(\pi,0)$;
however, as this point is only 20 meV below the Fermi energy, little
information is available before the band disperses through the Fermi energy.
For BSCCO, the tight-binding fit was done to the proposed band structure
presented by Dessau {\it et al.} based on their photoemission data and
yields a band width of 1.4 eV. \cite{Shen}
The dispersion was fit so as to go like $k^4$ along both orthogonal
directions about the $(\pi,0)$ point.
For comparison, we also include calculations for a flat density of states
with a 2 eV band width.

In computing the density of states from these energy dispersions, we employ
a two-dimensional ``tetrahedron'' code that produces data on an energy grid
of spacing 0.1 meV near the singularity.
We have checked that our
results are insensitive to this mesh density
and have reproduced our earlier calculations on a logarithmically
divergent density of states where the singularity
is treated exactly. \cite{Radtke}
We find that, for the LSCO and bifurcated YBCO cases, the density of states
has a logarithmic singularity.
For the extended YBCO case, on the other hand, the divergence goes like the
inverse fourth root of the energy difference from the singularity,
and the divergence in BSCCO goes as the inverse square root.
We remark that the densities of states are similar in both YBCO
cases, even though the divergences are different.
We see that, by studying these different materials, we are actually studying
different Van Hove singularity strengths.

The choice of coupling constant in our calculations is complicated due
to the sharp structure in the density of states.
In a material with a structureless density of states, the coupling
strength is $\lambda = 2 g^2 / W \Omega_0$.
However, the magnitude of the coupling extracted from transport measurements
will be renormalized by singularities in the density of states, and so
a more physical choice of coupling is to use the
self-consistently calculated self-energy at the lowest Matsubara
frequency $\omega_0 = \pi T_c$:
$\lambda_Z = - {\rm Im} \, \Sigma (i\omega_0) / \omega_0$.
In what follows, we will show calculations with either $\lambda$ or $\lambda_Z$
fixed and examine the differences.

For completeness, we will discuss some caveats regarding our approach.
First, our calculations employ a mean-field theory where vertex corrections
are ignored.
Although justified by Migdal's Theorem \cite{Migdal} in conventional
phonon-mediated superconductors with flat
densities of states, this approximation may fail when the density of
states is singular.
Recent work has indicated that, for a logarithmic
Van Hove singularity, vertex corrections may
not be large and may actually reduce the computed $T_c$, \cite{Kris}
but the case of a power law singularity has not been examined.
Second, it has been pointed out that, because the electronic density
of states enters into the screening of the electron-phonon matrix
element, ignoring the detailed wave vector dependence of this matrix
element and the phonon propagator--as we do--could lead to erroneous results.
\cite{Mahan,AAA}
Third, phase-space restrictions on electron-electron scattering
imposed by the Van Hove singularity enhances the electronic response
functions, possibly leading to an electronic pairing mechanism. \cite{Newns}
Our calculations focus on a phonon-like pairing boson
and so do not address this issue.
Fourth, for YBCO, we consider only the plane band which contains
the Van Hove singularity nearest the Fermi energy (for BSCCO, the splitting
of the two CuO plane bands is ignored).
We expect that this band will dominate the properties of the
material, but full multiband calculations will be required to test this
assertion.
We therefore view the results of this paper as a first step in identifying
the quantitative effect of a singular density of states on $T_c$
when strong-coupling effects are included and leave these other
extensions for future work.

Having described our approach, we will now discuss our results,
beginning with our weak-coupling calculations.
For these computations, we drop the
$\tau_0$ and $\tau_3$ components of the self energy from the Eliashberg
equations, and we fix $\lambda$ to 0.211 in order to obtain a
95 K $T_c$ for the YBCO band structure with the extended Van Hove singularity
and 2 eV band width.
Note that this weak-coupling calculation is not equivalent to
the standard BCS calculation with a square well potential, but does serve
to illustrate the effect of neglecting inelastic pair-breaking.

The maximum $T_c$ occurs at a band filling near but not necessarily at
the Van Hove singularity, and we tabulate these
maximum critical temperatures and isotope effect coefficients
as a function of material type in Tab. 3.
We observe that the maximum weak-coupling $T_c$'s are enhanced by factors
of 17 (LSCO) to 45 (BSCCO) over the flat density of states value and that
this enhancement follows the strength of the singularity, increasing
as the divergence goes from logarithmic for LSCO to
fourth root for the extended YBCO to square root for BSCCO.
In contrast, the isotope effect coefficient decreases systematically
with singularity strength from the
BCS result of 0.50 for the flat density of states to 0.20 for BSCCO.
We also see that the extended and bifurcated densities of states for YBCO
yield approximately equal $T_c$'s, despite the differences in singular
behavior.  Reducing the band width by one half, though, reduces $T_c$ by 40\%,
despite working at fixed $\lambda$.
The isotope effect coefficient, however, is unaffected by this
change.
Finally, note that the critical temperature roughly doubles from
the logarithmically divergent LSCO model to the fourth-root divergent extended
YBCO model, similar to the $T_c$'s of the actual materials.
This observation leads to the speculation that the pairing interaction is
the same in LSCO and YBCO with the difference in critical temperatures
accounted for solely by the band structure.

The strong-coupling calculations do not support this simple view, however.
In Tab. 3, we show strong-coupling results for the maximum $T_c$
obtained from the solution of the Eliashberg equations
including all components of the self-energy and with $\lambda$ = 0.627,
chosen once again in order to fix the $T_c$ for the W = 2 eV extended
YBCO band structure to 95 K.
We see that the critical temperatures still show a systematic increase with
the strength of the singularity, but the strong-coupling $T_c$'s show
less sensitivity to the singularity strength than the weak-coupling $T_c$'s:
the maximum strong-coupling $T_c$ is enhanced by factors of only
2.6 (LSCO) to 3.4 (BSCCO) relative to the flat density of states $T_c$.
Similarly, the strong-coupling isotope effect coefficient still exhibits
a systematic decrease with singularity strength, but the magnitude of this
trend is severely reduced, leaving $\alpha$ roughly constant at 0.4.
We also note that the finite band width and strong-coupling effects combine
to increase the flat density of states $\alpha$ to 0.51 instead of 0.50.
As with the weak-coupling calculations, the extended and bifurcated densities
of states yield similar $T_c$'s, $\alpha$'s, and $\lambda_Z$'s; reducing the
band width to 1 eV reduces $T_c$ and $\lambda_Z$ by approximately 20\% but
leaves $\alpha$ almost unchanged.

Drawing these results together, we can say that, in both weak- and
strong-coupling calculations, $T_c$ and $\alpha$ show systematic trends
with the strength of the divergence in the density of states,
but the magnitude of this effect is
reduced in the strong-coupling calculations.
In particular, we find that the isotope effect coefficient in strong
coupling is almost
independent of the strength of the singularity and is around 0.4, which
is much larger than the observed $\alpha$'s in the optimally doped
cuprates.
Also, the enhancement of $T_c$ relative to the flat density of states in
strong-coupling is two to three for our choice of $\lambda$,
as shown earlier for a logarithmically
singular density of states. \cite{Radtke}
Furthermore, we find that these results are robust against the parameterization
of the YBCO data (extended vs. bifurcated),
but $T_c$ and $\lambda_Z$ are found to be sensitive to band width (although
$\alpha$ is not).

Additionally, the idea of a universal pairing interaction does not seem to
be supported by the strong-coupling calculations.
By fixing the coupling in the extended parameterization of YBCO to give
a strong-coupling $T_c$ of 95 K, we find that the maximum $T_c$ for LSCO
becomes 78 K,
much larger than the observed maximum $T_c$ of 40 K.  This result
casts doubt on the idea that the differences in $T_c$ are simply due to
band structure effects.  We should also remark that the experimental
$T_c$ for BSCCO is lower than YBCO, despite the more singular density of
states inferred for the former.  We show a plot of $T_c$ versus
doping for this $\lambda$ in Fig. 2, which illustrates in greater detail
both the reduced difference in $T_c$ and the enhancement of
$T_c$ over that given by a flat density of states.

There are several other features of Fig. 2 which should be pointed out.
First, we note that the critical temperature rises around the van Hove
singularity, and the width of this maximum
broadens as the strength of the singularity increases.
Away from the singularity, the $T_c$ falls below the flat density of
states value.
This behavior is a reflection of the redistribution of electronic states
caused by the Van Hove singularity; basically, spectral weight is transferred
from the band edge to the singularity, and $T_c$ follows this shift.
Second, at low values of $n$, the plotted data show somewhat
more complicated behavior due to the effect of the band edge.
Finally, we see that the extended and bifurcated YBCO give
similar $T_c$ vs. doping curves despite the differing singular behavior.
The density of states peak in the extended case is higher than the bifurcated
case but is also narrower in energy.
This result supports the idea that what determines $T_c$ is an effective
density of states which is a particular average of the actual density of
states over the scale of the phonon energy $\Omega_0$.\cite{CarbotteSC,Pickett}

In looking at the self-consistent coupling strength $\lambda_Z$ in Tab. 3,
we see that $\lambda_Z$ is larger than $\lambda$ for the singular densities of
states by a factor of order two, implying that these calculations are
in the strong-coupling regime.
The systematic behavior of $\lambda_Z$ with $\lambda$ is shown in Fig. 3.
For a flat density of states and an infinite band width,
$\lambda_Z$ = $\lambda$.  For the finite band widths used here,
this relation
is valid only in the limit of small $\lambda$.  As seen in the inset to
Fig. 3, at larger
$\lambda$, the flat density of states $\lambda_Z$ is less than $\lambda$ by
about
a factor of two.  On the other hand,
when the density of states is singular, $\lambda_Z$ is enhanced at small
$\lambda$, and the magnitude of the enhancement grows as $\lambda$ decreases
and as the strength of the singularity increases.
This behavior arises from the availability of more scattering states near
the Van Hove singularity.
At large $\lambda$, however, the structure in the interacting density of
states is smeared out by the strong scattering, and $\lambda_Z$ for
the singular density of states begins to behave more like that for a
flat density of states; i.e., it is dominated by the finite band width.
The curves come together at large $\lambda$ as shown in the inset to Fig. 3.

Since both $T_c$ and $\lambda_Z$ are enhanced by the Van Hove singularity,
plotting the one quantity as a function of the other is instructive.
To that end, we show in Fig. 4 the maximum transition temperature
$T_c^{max}$ normalized by the Einstein phonon frequency $\Omega_0$
plotted against the self-consistent $\lambda_Z$ for all the materials
studied and with the results for a flat density of states included
for comparison.
Our results for the flat density of states
reproduce the standard curve at small $\lambda_Z$. \cite{AD}
Also at small $\lambda_Z$, $T_c$ is still strongly and systematically enhanced
by an increasingly singular density of states compared to the results for
the structureless density of states.
This enhancement is reduced at stronger coupling, becoming at most around 50\%
for $\lambda_Z \approx 1$.
At still stronger coupling, as seen in the inset to Fig. 4, the effect of
the singularity in the density of states is completely removed and the
critical temperature follows the results for a structureless density of
states.
If $T_c^{max} / \Omega_0$ is plotted against the bare
$\lambda$, the qualitative features of the curve are the same, but the
$T_c$ enhancement is larger as one would expect from Tab. 3 and the results
at very strong coupling do not scale as well as when the $T_c$'s are plotted
against $\lambda_Z$.

We can understand these results in the following way.
At small $\lambda$, $\lambda_Z$ is strongly enhanced by the presence of
a singularity in the density of states, but it is approximately equal to
$\lambda$ if there is no such structure.
Comparison of critical temperatures at fixed $\lambda_Z$ therefore
means that a much larger
value of $\lambda$ is used in the flat density of states calculation
and so the apparent effect of the singular density of states is reduced.
At strong coupling, the enhanced scattering from the singular bare density
of states acts to smear out the interacting
density of states and so there is no $T_c$ enhancement.
We note that the loss of the Van Hove singularity in the interacting
density of states at strong coupling has been seen explicitly in recent
calculations. \cite{Zhong}

It is clear from our results
that strong-coupling calculations are required to make
physical predictions of the critical temperature.
The remaining question is:  should one compare these strong-coupling
critical temperatures at fixed $\lambda$ or fixed $\lambda_Z$?
In our investigation of the possibility of a universal pairing interaction,
it is the microscopic coupling $g$ and phonon energy $\Omega_0$ which
would be fixed, yielding--for a given band width--a fixed $\lambda$.
Our results in this case show little evidence for a universal pairing
interaction
due to the reduction of the effect of the Van Hove
singularity on $T_c$.
Alternatively, one can extract an effective coupling constant from
transport measurements which would include the self-consistent effects
of the density of states and so could be used to fix $\lambda_Z$.
{}From earlier work,\cite{RULN,Radtke} we found that the canonical values of
$\lambda_{tr} \approx$ 0.2 to 0.4 extracted from optical conductivity
in YBCO
correspond to $\lambda_Z \approx$ 1, at least if spin fluctuations are
the source of the resistivity.
Similar estimates of $\lambda_Z$ have appeared in the literature regarding
phonon coupling strengths. \cite{Allen}
Based on this result and Fig. 4, we conclude that, while the Van Hove
singularity does enhance $T_c$ systematically with the strength of the
divergence in the density of states, the enhancement factor in this
parameter region is at most around 50\% and so by itself cannot account
for the large $T_c$'s observed.
In order to obtain a large $T_c$, either a large pairing interaction energy
$\Omega_0$ or a modification of the pairing interaction due
to the singularity is required, but it not known whether either effect
is consistent with the observed linear resistivity of the cuprates.

In conclusion, we have performed Eliashberg computations
of the critical temperature in superconductors with strongly singular
densities of states chosen to model LSCO, YBCO, and BSCCO.
We have demonstrated that, while the weak-coupling solutions do suggest
a strong influence of the strength of the Van Hove singularity on $T_c$
and the existence of a universal pairing interaction in the cuprates,
strong-coupling calculations show a reduced sensitivity to the singularity
strength and do not support the notion that band structure effects alone can
account for either the large $T_c$'s or the different $T_c$'s within the
family of copper oxide superconductors.
The reduced effect of the Van Hove singularities can be traced to the fact
that the inelastic scattering included in strong-coupling calculations
smears out the sharp structure in the density of states, producing a $T_c$
similar to that from a flat density of states.
This effect is seen clearly when the results are plotted as a
function of the physically relevant self-consistent coupling constant
$\lambda_Z$ and leads to universal behavior at very strong coupling.
If the Van Hove singularities in the electronic band structures of the
cuprates are responsible for high-temperature superconductivity,
then the effect must come in through vertex corrections, screening
effects in the electron-pairing boson matrix elements, direct electronic
interactions, or multi-band effects.
Based on the results of this work, investigation into these effects now
acquires additional importance.

\acknowledgements

This work was supported by the National Science Foundation (DMR 91-20000)
through the Science and Technology Center for Superconductivity
and DMR-MRL-8819860 (R.J.R.) and
by the U.~S.~Department of Energy, Basic Energy Sciences,
under Contract \#W-31-109-ENG-38 (M.R.N.).

\begin{figure}
\caption{Electronic energy dispersion along high-symmetry directions
in the Brillouin zone resulting from fits of (a) an LDA band structure
calculation for LSCO and (b) ARPES measurements on
BSCCO \protect\cite{Shen} to the
tight-binding form described in the text.
The wave vectors in the two-dimensional Brillouin zone
are $\Gamma = (0,0)$,
X = $(\pi,\pi)$, and G (M) = $(\pi,0)$.}
\label{fig:disp}
\end{figure}

\begin{figure}
\caption{Critical temperature $T_c$ in K as a function of
the number of electrons per unit cell away from the optimum value $n - n_{max}$
computed for the
tight-binding band structures corresponding to BSCCO (short-dashed line),
bifurcated YBCO (long-dashed line), extended YBCO with 2 eV band width
(solid line), and LSCO (dot-dashed line).
The horizontal solid line is the value of $T_c$ for the flat density
of states at half-filling.
These results are produced by strong-coupling
Eliashberg theory with an Einstein phonon of energy $\Omega_0$ = 500 K and
$\lambda$ = 0.627.
See Tab. 3 for additional results at optimal doping $n_{max}$.}
\label{fig:doping}
\end{figure}

\begin{figure}
\caption{Self-consistent coupling constant $\lambda_Z$ as a function
of the bare coupling constant $\lambda$
for the band structures corresponding to LSCO (solid boxes),
YBCO with the extended (open diamonds) and bifurcated (open triangles)
Van Hove singularities and 2 eV band widths,
BSCCO (solid triangles), and a flat density of
states of 2 eV band width (open boxes).
The lines are to guide the eye.
The $\lambda_Z$'s result from the solution of the
strong-coupling Eliashberg equations as described in the text.
Inset:  Data from the main figure displayed out to large $\lambda$.}
\label{fig:lambdaZ}
\end{figure}

\begin{figure}
\caption{Maximum critical temperature $T_c^{max}$ normalized by
the Einstein phonon
frequency $\Omega_0$ as a function of the self-consistently determined
$\lambda_Z$.
Symbols are the same as in Fig. \protect\ref{fig:lambdaZ}, as is the method
of computation.
Inset:  Data from the main figure displayed out to large $\lambda_Z$
in order to demonstrate the absence of density of states effects at
very strong coupling.}
\label{fig:normTc}
\end{figure}

\narrowtext
\begin{table}
\caption{Two-dimensional, tight-binding basis functions used in
the fits of the energy dispersion described in the text in
units of the lattice spacing (a square unit cell is assumed). }
\begin{tabular}{cc}
$i$ & $\eta_i ({\bf k})$ \\
\tableline
0 & $1$ \\
1 & $\frac{1}{2} (\cos k_x + \cos k_y)$ \\
2 & $\cos k_x \cos k_y $ \\
3 & $\frac{1}{2} (\cos 2 k_x + \cos 2 k_y)$ \\
4 & $\frac{1}{2} (\cos 2k_x \cos k_y + \cos k_x \cos 2k_y)$ \\
5 & $\cos 2k_x \cos 2k_y $ \\
\end{tabular}
\end{table}

\begin{table}
\caption{Parameters $t_i$ in eV corresponding
to the basis functions in Tab. 1 used to fit the LDA calculations and
photoemission measurements of several cuprate superconductors.
When $t_i$ is zero, this parameter was not included in the fit.  For YBCO,
in parenthesis are the band width (eV) and saddle point type (ext for
extended, bf for bifurcated).}
\begin{tabular}{cccccc}
  & LSCO & YBCO(1,ext) & YBCO(2,ext) & YBCO(2,bf) & BSCCO \\
\tableline
$t_0$ &\dec  0.00 &\dec  0.00 &\dec  0.00 &\dec  0.00 &\dec  0.0607 \\
$t_1$ &\dec -1.00 &\dec -0.50 &\dec -1.00 &\dec -1.00 &\dec -0.525 \\
$t_2$ &\dec  0.20 &\dec  0.15 &\dec  0.38 &\dec  0.38 &\dec  0.100 \\
$t_3$ &\dec  0.00 &\dec -0.05 &\dec -0.06 &\dec -0.09 &\dec  0.0287 \\
$t_4$ &\dec  0.00 &\dec  0.00 &\dec  0.00 &\dec  0.00 &\dec -0.175 \\
$t_5$ &\dec  0.00 &\dec  0.00 &\dec  0.00 &\dec  0.00 &\dec  0.0107 \\
\end{tabular}
\end{table}

\widetext
\begin{table}
\caption{Band filling corresponding to maximum critical temperature, $n_{max}$,
maximum critical temperature, $T_c^{max}$, isotope effect coefficient,
$\alpha$,
and self-consistent coupling constant, $\lambda_Z$, for the densities of
states discussed in the text.  Results are obtained from weak-coupling
Eliashberg theory with $\lambda$ = 0.211 and strong-coupling Eliashberg
theory with $\lambda$ = 0.627 in order to give a 95 K $T_c$ in the extended
YBCO band structure with 2 eV band width.  YBCO notation as in Tab. 2.}
\begin{tabular}{ccccccc}
  & Flat & LSCO & YBCO(1,ext) & YBCO(2,ext) & YBCO(2,bf) & BSCCO \\
\tableline
$n_{max}$   &\dec 1.00 &\dec  0.82 &\dec 0.55 &\dec 0.48
     &\dec  0.45 &\dec  0.92 \\
Weak-coupling \\
$T_c^{max}$ (K)   &\dec 3.1 &\dec 54.2 &\dec 59.4 &\dec  95.0
     &\dec 97.5 &\dec  139. \\
$\alpha$    &\dec 0.50 &\dec  0.34 &\dec 0.28 &\dec  0.28
     &\dec  0.27 &\dec  0.20 \\
Strong-coupling \\
$T_c^{max}$ (K)   &\dec 30.5 &\dec  78.0 &\dec 77.9 &\dec  95.0
     &\dec  94.9 &\dec  105. \\
$\alpha$    &\dec 0.51 &\dec  0.43 &\dec 0.40 &\dec  0.41
     &\dec  0.41 &\dec  0.40 \\
$\lambda_Z$ &\dec 0.59 &\dec  1.07 &\dec 0.96 &\dec 1.30
     &\dec  1.29 &\dec  1.45 \\
\end{tabular}
\end{table}

\end{document}